# Full-privacy secured search engine empowered by efficient genome-mapping algorithms

Yuan-Yu Chang[1,2]†, Sheng-Tang Wong[2]†, Emmanuel O Salawu[1,3,4]†, Yu-Xuan Wang[2], Jui-Hung Hung[5]* and Lee-Wei Yang[1,3,6,7]*

[1]Institute of Bioinformatics and Structural Biology, National Tsing Hua University, Hsinchu 300044, Taiwan; [2]Praexisio Taiwan Inc. New Taipei 221425, Taiwan; [3]Bioinformatics Program, Institute of Information Sciences, Academia Sinica, Taipei 115201, Taiwan; [4]Machine Learning Solutions Lab, Amazon Web Services (AWS), Herndon, VA, USA; [5]Department of Computer Science, College of Computer Science, National Yang Ming Chiao Tung University, Hsinchu 300044, Taiwan; [6]Physics Division, National Center for Theoretical Sciences, Taipei 106216, Taiwan; [7]PhD program in Biomedical Artificial Intelligence, National Tsing Hua University, Hsinchu 300044, Taiwan.

† Authors who have equal contributions

* To whom correspondence should be addressed: Lee-Wei Yang. Tel: +88635742467; Fax +88635715934; Email: lwyang@life.nthu.edu.tw; Jui-Hung Hung. Email: juihunghung@gmail.com

## Abstract (250 words)

Search engine privacy has not been an option for web exploration in the past. Search history returns to users as personalized merchandise promotion from search engine companies. Here we demonstrate a privacy-guaranteed internet search method, provided with the entire article as a query, can be correctly carried out without revealing users' sensitive content by a degenerate, irreversible, encoding scheme and an efficient FM-index search routine, generally used in the next-generation sequencing. In our solution, Sapiens Aperio Veritas Engine (S.A.V.E.), every word in the query is encoded into one of 12 "amino acids" comprising a pseudo-biological sequence (PBS) at users' local machines. The PBS-mediated plagiarism detection is done by users' submission of locally encoded PBS through our cloud service to locate duplicates in the collected web contents, currently including all the English and Chinese Wikipedia pages and Open Access journal articles, as of April 2021, which had been encoded in the same way as the query. It is found that PBSs with a length longer than 12, comprising a combination of more than 12 "amino acids", can return correct results with a false positive rate <0.8%. S.A.V.E. runs at a similar mapping speed as Bowtie and is 5-order faster than BLAST. S.A.V.E., functioning in both regular and in-private search modes, provides a new option for efficient internet search and plagiarism detection in a compressed search space without leaking users' confidential contents. We expect that future privacy-aware search engines can reference the ideas proposed herein. S.A.V.E. is currently running at https://dyn.life.nthu.edu.tw/SAVE/

## INTRODUCTION

The worldwide web has been navigated and accessed by search engines since the middle of 90s. Since then, people used to "google" the web to locate the needed information instead of always bookmarking them. The indispensability of search engines comes with a price of compromised personal privacy where one's query histories, browsing habits, and other personal info are collected for improved search-result relevance, while this personalization also accompanies targeted ads delivery or sales of users' data to third parties (*1*). A study suggested that a user's query can be stored by the search engine company for 18 months (*2*). There are search engines that operate by storing less personal information (*3*), (*4*) or hiding users' IPs through thousands of relays (*5*). These search methods, however, have not offered guaranteed security and intrinsic privacy for users.

Free access to online information also brought reckless misuse and unauthorized copies from internet sources. Full-article-query-based (FAQ-based) search engines have been used as a plagiarism detection tool (*6*) since the beginning of this century. Plagiarism constitutes the re-use of other people's language, ideas, and expressions in one's work without proper acknowledgment (*7*), usually through a direct copy-and-paste (*8*), which surely undermines educational efforts and originality of scientific researches (*8, 9*). To mitigate such an issue, a number of algorithms and software have been developed, where pre-built databases and the use of search engine APIs are two primary searching strategies. The former builds databases in advance using academic journals, magazines, news, blogs, and users' articles, for plagiarism detection. For instance, Turnitin (*10*), Grammarly, Duplichecker and eTBLAST (*11*) operate with pre-built databases. On the other hand, search-engine-based plagiarism detection tools, such as Plagiarisma and SNITCH (*12*), directly query search engines with the content of users' articles instead of building a local database. Both types of applications require users to upload their documents to specified remote servers. The queries are compared with what is stored in the database or with results returned from popular search engines. With both, users always have limited control of their submitted content; data leakage can happen during transmission or under malicious attacks. Also, service providers could reuse uploaded articles for expanding their current databases for profit (*13*) or for commercial uses.

In order to provide a search service with 100% privacy protection of a given query, either for expertise matching (*11*) or plagiarism detection, we have developed the Sapiens Aperio Veritas Engine (S.A.V.E.), as a plagiarism detection solution conducting plagiarism detection with full protection of user-uploaded content. S.A.V.E. takes irreversible encryption of query articles or strings for a fast database search. The irreversible encryption mapped each word or character from an article into an "amino acid", which is a single character (see **Figure 3** for example). The pseudo-biological sequences (PBS) are composed of sets of aforementioned encrypted characters, which is analogous to biological sequences, where a long article can be a pseudo "genome". The search for this genome can therefore lend itself to any state-of-the-art genome mapping algorithm - in our case, the FM-indexing. FM-index was first introduced in 2000 (*14*) as an opportunistic data structure based on the Burrows-Wheeler transform (BWT) (*15*) that compresses the input text for subsequent efficient substring search. The advantages of the FM-index make it suitable for DNA read mappings, which led to the rise of multiple alignment tools based on FM-index, such as Bowtie (*16*),

BWA (*17*), and SOAP2 (*18*). The FM-index-based approaches allow an efficient query, which takes no more than 5 seconds for S.A.V.E. to detect plagiarism among English and Chinese Wikipedia pages, PubMed abstracts and open access articles provided with an article with thousands of words. Given the irreversible encryption (ciphertext), the exact match could refer to more than one version of the plaintext, producing false positives. In this report, we minimized the chance of obtaining false positives by studying the sensitivity of queried PBSs as a function of PBSs in different lengths comprising different numbers of constituent amino acids (see below). We also found our implementation enjoys a similar running speed as Bowtie's while being >500,000 times faster than BLAST.

## RESULTS

### Plagiarism detection framework

The plagiarism detection framework of S.A.V.E. provides the highest privacy protection by not revealing the original content of the query document. S.A.V.E. uses destructive encryption to encrypt a document into a long pseudo-biological sequence (PBS) comprising a fixed number of characters, as an irreversible ciphertext. For instance, in our current implementation, any to-be-uploaded document(s), written in any language, can be translated into a long, irreversible cyphertext comprising 12 characters (i.e., A, C, D, E, G, H, I, K, L, N, Q and R, where $a = 12$) such as "NAGNDDHHIGGQLCKNNNAAQKKKDQARDDCIGLLNDANCQQNRNRRDEQRAREA..." by our local software in users' personal computer before being uploaded (to S.A.V.E.). S.A.V.E. takes both plaintext and cyphertext where the former is not discussed in this article.

In the latter scheme, S.A.V.E. separates the encryption function from the searching function to prevent the exposure of users' sensitive content, where the encryption is performed with a local device, and the searching is performed in the cloud. As shown in **Figure 1**, users can encode their documents to query by an offline software we provide, which can be downloaded from https://dyn.life.nthu.edu.tw/SAVE/download. Next, users upload the encrypted documents to S.A.V.E. for plagiarism detection. The degenerate encoding nature makes data stolen, either in transmission or on the server, impossible. In the S.A.V.E. database (top of the grey area in **Figure 1B**), web content is encoded in the same way as by the local software. The search and detection are carried out by a modified version of FM-index used in next generation sequencing (see Methods). After the search is finished, users can download the metadata of the search results in JSON format and generate the report with S.A.V.E. local software alongside the original documents. After the JSON-based meta-files are downloaded, all the operations to decipher the plagiarized loci in plain text can be done offline. At the server end, two different pathways of plagiarism detection are implemented depending on the number of uploaded documents. When users upload a single encrypted document, S.A.V.E. searches its PBS database containing encrypted web content that currently includes all the English and Mandarin Wikipedia pages and open-access journal articles (**Figure 1B**). If users zip multiple encrypted documents into a single file and then upload the zip file to S.A.V.E., S.A.V.E. would first build a temporary database based on these uploaded PBSs and then check their mutual plagiarism in a one-to-multiple manner (also called "pairwise comparison", **Figure 1A**). After the pairwise comparison step, users can also select all or a subset of the documents for the search against encrypted web content using the pre-built S.A.V.E. database as what is described in Figure 1B (almost the same as Figure 1A but is for a single document). Both pathways result in a metafile (in Python JSON format) to be downloaded into a local folder in users' local computers, which contains local mapping files and the original documents. The S.A.V.E.-returned metafile contains the mapping information where stretches in a PBS match with the corresponding texts in the web content, while the local mapping files have the location mapping information between users' original plaintext and the corresponding PBS encryption. As a result, when putting together both of these files, our local software can resolve pieces in the original plaintext that are duplicated from specific locations in given web content, presented with a local web page that is in the same format as what users can find in our S.A.V.E. website.

With partitioning the encryption done locally and the searching conducted in the cloud, S.A.V.E. can protect users' intellectual properties while executing plagiarism detection with quality and efficiency (see below).

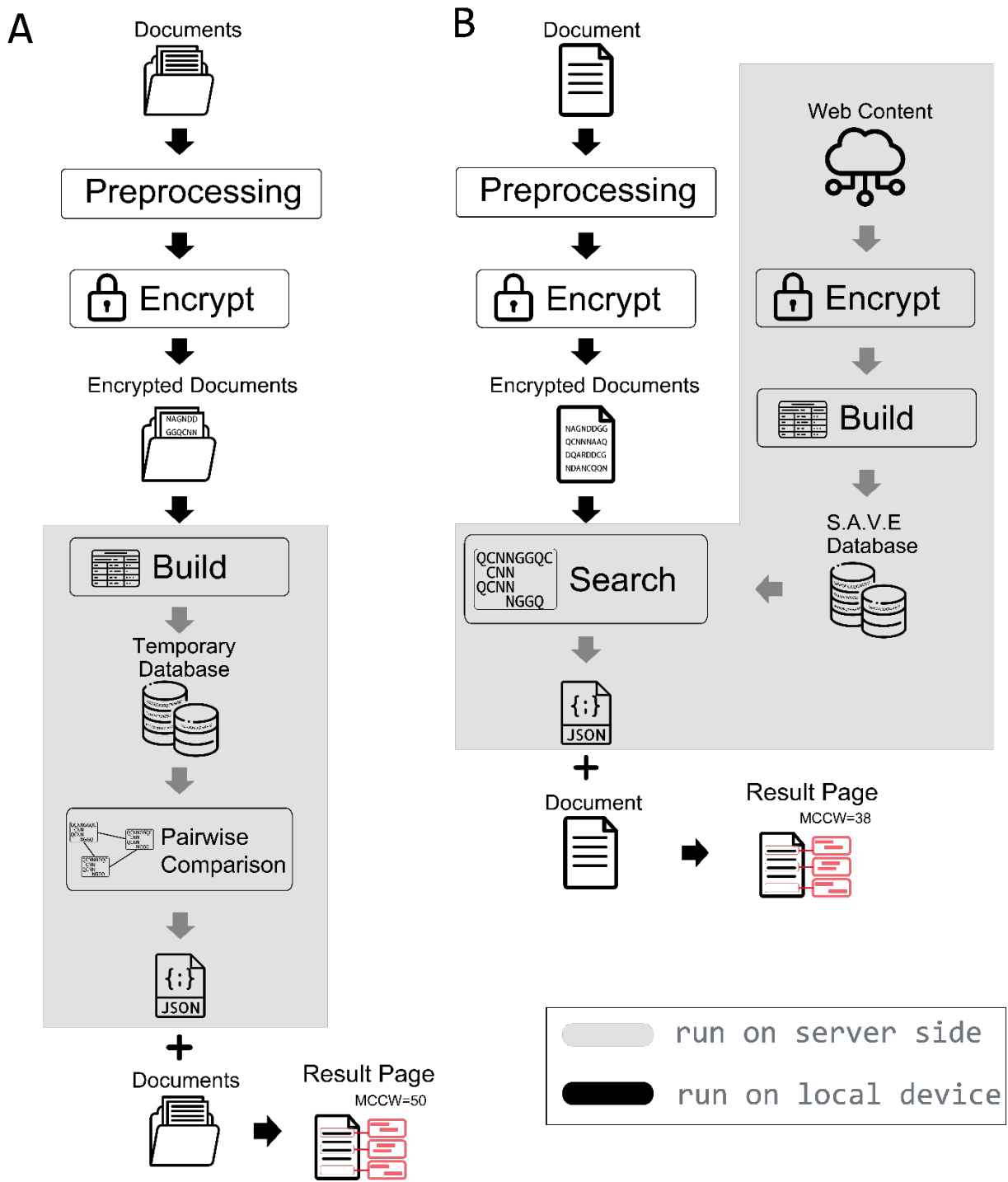

**Figure 1**. The plagiarism detection framework. **A.** The pairwise comparison for multiple documents. Users can encrypt their confidential documents on their local computers before uploading them to the S.A.V.E. server; by doing this, the sensitive content of the documents is not revealed to search engine providers and is free from the risk of being intercepted during the transmission. On the server-side (in the grey area), the uploaded encrypted files are first used to build a temporary database followed by a pairwise comparison to generate a JSON file. With the JSON metadata and the original file stored on the user's local device, S.A.V.E. allows users to browse the result web page, highlighting the locations in the original documents copied from each other, in their local device (**Figure 2A, bottom**). **B.** PBS encoded from a single document. In this pathway, S.A.V.E. detects plagiarism of the document against pre-compiled PBSs

encoded from web content (supposedly the whole web) instead of performing the pairwise comparison. Users can browse their results directly on the S.A.V.E. website (**Figure 2B**).

**Assessment of the sensitivity and false positives**

Since S.A.V.E. uses irreversible encryption for the plaintext, which is a multiple-to-one mapping, two or more pieces of plaintexts can correspond to the same PBS. This is to say that in-private searches using PBSs may return us with results where the discovered plaintexts are not the same as those originally encoded. To understand at what false positive rate our search algorithm has as a function of different lengths ($k$ = 8, 10, 12, 14 or 16) of queries comprising a selection of $a$ ($a$ = 8, 12, 14 or 16) characters ("amino acids"), we conducted the following assessment.

To evaluate the false positive rate, the chance to find a piece of plaintext in the database that is not the original plaintext, encoded into the query PBS, we investigate how likely this happens when the search starts from a $k$-mer PBS, where $k$ = 8, 10, 12, 14 or 16. Note that in each position of a $k$-mer PBS stands one of $a$ ($a$ = 8, 12, 14 or 16) characters. Since a $k$-mer PBS always corresponds to a piece of plaintext comprising $k$ (human) words, we transform the problem into a space spanned by all the $k$-word segments in the Chinese and English Wikipedia pages and then examine the chance that different $k$-word segments correspond to the same PBS (as if our users search an encoded $k$-word plaintext and then ask what the chance for our S.A.V.E. engine to return results containing plaintexts that are not the originally encoded plaintext). We consider only unique $k$-word strings to avoid the double counting. In 9,266,370,827 unique 12-word strings ($k$=12; "unique" means that a given 12-word segment only appears once in all the English and Chinese Wikipedia pages) when $a$=12, there are 70,352,323 such strings encoded into a PBS that other strings are also encoded into, hence a false positive rate of 70,352,323 / 9,266,370,827 = 0.0076. Results of other $k$-word strings can be found in **Table 1**.

**Table 1 False positive rate among various parameters**

| Length of the k-mer string | false positive rate (%) | | | |
| | number of characters used ($a$) | | | |
| | 8 | 12 | 14 | 16 |
|---|---|---|---|---|
| 8 | 100.00 | 100.00 | 99.65 | 87.87 |
| 10 | 78.71 | 16.28 | 4.39 | 1.99 |
| 12 | 4.49 | 0.76 | 0.55 | 0.58 |
| 14 | 1.26 | 0.49 | 0.43 | 0.46 |
| 16 | 1.02 | 0.40 | 0.36 | 0.37 |

It can be found that length plays a big role in determining the false-positive rates. When the length is 8, the false positive rate can be as high as 87.87% even if 16 characters ($a$ = 16) are used.  As the length of PBS grows, length 10 with 16 characters can already reduce the false-positive rate to <2.00%. No larger than 5.00% false positive rate can be found for the PBSs longer than 12, while 12-mer PBSs with $a$ = 12 characters have a small false-positive rate <0.8%. To consider detection sensitivity (the shorter $k$ the better), high security due to encoding degeneracy (the smaller $a$ the better) and false-positive rates (the larger the k and $a$ are, the better), we therefore chose $a$ = 12 and $k$ = 12 as our encoding and sensitivity scheme for S.A.V.E. Duplicates less than $k$ = 12 are not reported by our software and long stretches of duplicated texts are detected as overlaps of many 12-mer PBSs.

**The encoding scheme reduces the size of S.A.V.E. databases**

The destructive encryption algorithm of S.A.V.E. is not only irreversible but also compressed, where a word (e.g., Utopia, CRISPR-Cas9, 新 or は) is encoded into one alphabet which certainly reduces the size of a document tremendously, although the application of the herein introduced FM-index alignment algorithm for plagiarism detection requires extra reference tables (see **Figure 4**), leading to a slight increase (less than 1.8-fold) in the size (**Table 2**). Table 2 shows that our FM-indexed PBSs with needed reference tables are about 1/6 of the size of original web content (146.1 GB, considering the text only).  The encryption of words for phonemic western languages (e.g., English, French or Spanish etc.) resulted in a bigger compression ratio (e.g., 5.9 for English Wiki pages), while the ratio for non-alphabetic eastern language (e.g., Chinese, Korean, Japanese Kanji or Tamil etc.) is understandably smaller (e.g., 2.1 for Chinese Wiki pages) because one eastern "character" (e.g., 字) is one word.

**Table 2. Compression ratios of our encoding for different texts**

| Database | Capacity (GB) | | | compression ratio* |
|---|---|---|---|---|
| | Raw | PBS | FM-Index | |
| Pubmed OA Commercial[†] | 58.7 | 5.9 | 9.8 | 6.0 |
| Pubmed OA Noncommercial[†] | 24.2 | 2.4 | 4.0 | 6.1 |
| Pubmed Baseline | 45.6 | 4.6 | 7.4 | 6.2 |
| Wiki English | 16.0 | 1.7 | 2.7 | 5.9 |
| Wiki Chinese | 1.6 | 0.5 | 0.8 | 2.1 |
| Total | 146.1 | 15.1 | 24.7 | 5.9 |

\* Compression Ratio = size of raw data / size of the index;

**Logistic regression model to remove references in the plaintext**

A bibliography or list of references is the source of false positives either in plaintext or when encoded into PBSs. This is because a listed reference is definitely the same or very similar to what can be found online. In a normal report, journal article or thesis, references are listed to acknowledge the sources of our background introduction, employed methodologies and arguments that corroborate our results. These references in user-uploaded articles, encoded into PBSs, should not be considered part of plagiarism, and therefore their removal prior to the PBS-encoding and upload is essential to reduce false positives. Since the running environment may be in users' personal computers, the compatibility and portability of the reference detector are important. We used a logistic regression model to identify the lines in the article that belong to the reference and those that are not. The training and testing sets contain 100 journal papers (containing ideal, well-formatted references) and 200 student reports (containing unideal and often poorly formatted references), where symbols such as ";", "-", "(", ")" etc. and four-digit numbers, representing years, are used as training features (see details in METHODS). The performance of the trained model reaches 0.97 area under the curve (AUC) of the receiver operating characteristic (ROC) curve, where unity is the highest possible value and 0.5 is the lowest. It is worth noting that the reference "lines" do not have to be at the end of the article for correct detection, considering the situation where figures or supporting information can be listed after the references, since reference position is not one of the training features in the logistic model.

**Speed analysis**

We compared the execution time of our FM-index implementation with Bowtie, a popular FM-index-based NGS aligner, and BLAST (*19*), a classic bioinformatics aligner. FM-index is the most popular algorithm in bioinformatics and is the most efficient algorithm in string matching, of which the time complexity is $O(m)$, where $m$ is the length of queries.

To fairly assess these tools, we generated genomes and queries of the same length, which are 500 million bases and 1 million 12-base long reads (the queries), respectively. The genomes and the queries for our implementation comprise 12 characters (i.e., $a$=12, comprising A, C, D, E, G, H, I, K, L, N, Q and R) while for Bowtie and BLAST the queries comprise 4 characters (i.e., $a$=4, comprising A, C, G and T). Each base of the genomes is selected randomly from the character set and each of the queries is sampled indiscriminately from the corresponding genome.

The results showed that the throughput of our implementation was 147.6 million bases per second, Bowtie was 73.33 million bases per second, and BLAST was 291 bases per second, which is about a half-million times slower than our current implementation. However, after the FM-index search, Bowtie performed local alignment that our implementation did not, which partly explains why Bowtie's speed was twice lower than ours. The efficiency of our implementation was a half-million times better than BLAST, as one would expect - FM-index is intrinsically faster than non-FM-index algorithm.

## Local Application, S.A.V.E. database and interfaces

To use S.A.V.E. to detect plagiarism in the privacy-protective (private) mode, one needs to download the local software from https://dyn.life.nthu.edu.tw/SAVE/download, which allows one to encrypt her/his document into a PBS for the subsequent upload and detection. When multiple files need to be detected, users can first zip the documents into a single zip file and our local software would then unzip the file, extract the text from each document, convert each document into a file containing the corresponding PBS, and then zips the resulting PBSs back into a zipped file again for subsequent upload and detection (see Introduction videos in https://dyn.life.nthu.edu.tw/SAVE/static/video/video.html ). Alternatively, one can also copy the PBS encoded from a document and paste it into the search bar of S.A.V.E.'s front page.

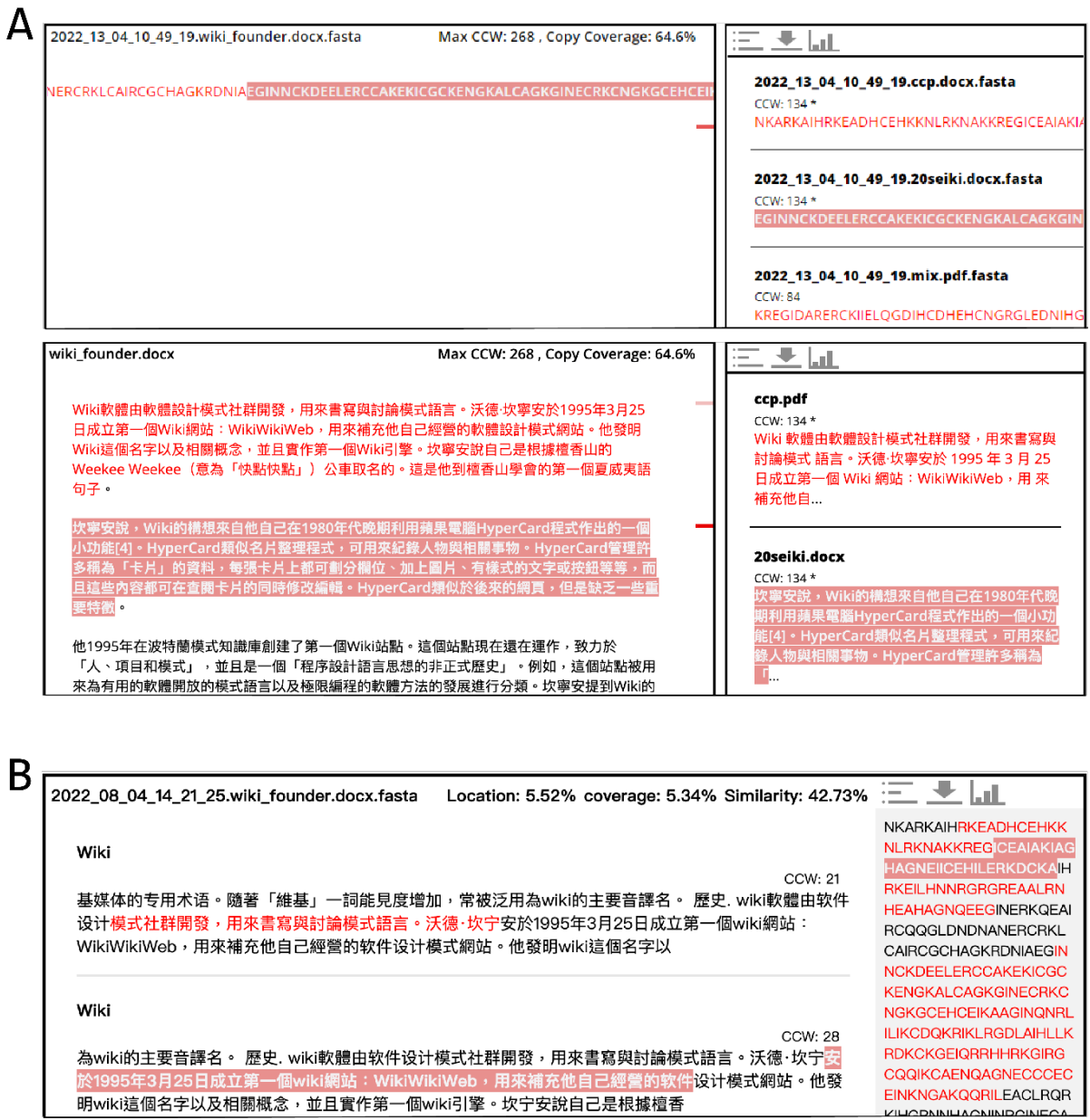

**Figure 2.** The web interface showing plagiarism detection results with a PBS query. **A. (top)** For the pairwise comparison of a batch of uploaded PBSs, results demonstrated on the S.A.V.E. website are (1) which two files have mutual plagiarism and (2) the locations of duplicated PBS segments in the two files. In the demonstrated case, founder.docx has chunks of texts (dynamically highlighted in red) being word-by-word copied into or taken from the three documents listed on the right where the locations of the duplicates are dynamically highlighted; **(bottom)** The pairwise comparison results are rendered in users' local devices after the JSON metadata (Figure 1A) are downloaded from the S.A.V.E. website to where users' original plaintext files are stored. The shown webpage is rendered by the S.A.V.E. desktop software. **B.** The internet search results show a user-uploaded document as the PBS on the right-hand side and the plaintext of copied sources from the internet on the left. Corresponding texts on both sides are simultaneously and dynamically highlighted when the cursor hovers on the relevant texts.

When a zipped file containing multiple PBS files is uploaded (**Figure 1A**), all the PBSs will first be pairwise compared to identify mutually plagiarized PBSs (files); in other words, S.A.V.E. server examines which PBS files contain identical PBS stretches, longer than 12 encoded alphabets in length. One report is returned for each PBS (file). In this html-formatted report, the PBS file of interest is present together with the files that plagiarize this file (share identical PBS stretches) as well as the location where plagiarism occurs (**Figure 2A, top**). At this stage, only PBSs are present, not all the corresponding plaintext that are not uploaded to the server in the first place. Currently, S.A.V.E. allows users to download the metadata, which can be used to map the original plaintext to plagiarized parts of the PBS through referencing locally stored original documents. Users' local folders have had the mapped location information between the original text and the PBS into which it is encoded, while the downloaded metafile contains the information on which parts of the uploaded PBS have the plagiarized sources in the plaintext format. With the metadata, users can render the report on their local devices (**Figure 2A, bottom**).

Also, after the pairwise comparison finished, all or selection of these PBS files can be marked for a database search to identify plagiarism from online sources. Detection result for each PBS query is presented in the same way as that obtained when a single PBS is being uploaded (**Figure 1B, 2B**) where plagiarized plaintexts in Wikipedia pages or Open Access journal articles are highlighted (left, **Figure 2B**) side by side with the corresponding PBSs in the query (right, **Figure 2B**). The length of the longest verbatim copy is reported along with how much percentage of the uploaded PBS can be found to duplicate the texts in Wikipedia pages and Open Access Journals.

Although major search engine companies and plagiarism detection services have policies to protect users' uploaded information, the promises are often compromised during data transmission or constant malicious attacks. Our design provides a prototype to fundamentally eliminate such a possibility.

Although it is beyond the scope of this article, S.A.V.E. does support the non-private search, conducting the plagiarism detection using one's uploaded plaintext(s). The functionality on this is detailed in the online user guide and the video tutorial at https://dyn.life.nthu.edu.tw/SAVE/.

## Discussions
### The new choice of a search engine that provides ultimate privacy protection of uploaded query texts
S.A.V.E. enabled plagiarism detection for confidential documents without revealing the exact content of the query, which made it impossible for data stolen during transmission or from the servers as well as storage of user-provided content for profit (*20*).

**A lower cost to maintain databases for plagiarism detection without compromising the copyright protection for subscription-based articles**

One of the biggest costs for plagiarism-detection is to maintain a database containing subscription-based journal articles and other online sources with copyright protections. Our current solution suggests a future possibility to store copyright-protected content in the form of encrypted PBSs. The detection results link back to the corresponding plagiarized sources from journal publishers, patent offices, government agencies, banks or hospitals without infringing their copyrights. In this setting, PBSs and their corresponding article URLs are stored. A URL reveals the full content of an article only when the viewers browse the content from an authorized IP address (by paying the subscription fee to the journal publisher), but reveals only the abstract for non-subscribers. In other words, S.A.V.E. stores the PBSs of web content, including those encoded from subscriber-based online articles, as well as their URLs. Users search S.A.V.E. database with query-encoded PBS and find the match. If the match has corresponding plaintext stored in S.A.V.E. database, a metafile can be downloaded for resolving the exact locations of the matched text; if not, a URL is provided to point the web source for subscribers (while non-subscribers cannot) to download the original piece. This can be then followed by uploading both the query and the downloaded original article to S.A.V.E., in either plaintext mode or encoded mode, for plagiarism detection in a pairwise comparison.

**Parallelized search in a compressed space**

The compressed nature of our encoding scheme also promises a reduced storage space (and therefore the cost), while our search scheme easily allows a search parallelization and database expansion. For instance, sBWT, using Schindler transform, is readily deployable using GPU-accelerated hardware solution (*21*). For the false positive issue, even if a longer stretch of PBS is used, the current implementation could have an intrinsic false positive rate of $1/a$ where $a$ is the number of constituents "amino acids" (see **Table 2**). It is inevitable that similar sentences can be encoded into identical PBSs. However, in such cases, to flag a potential plagiarism event with one or two words in a long string being actually different may not violate how most of us conceive plagiarism.

**More than one appropriate metric is included to quantify the extent of plagiarism**

Many plagiarism detection services report the plagiarism percentage in uploaded articles, which *de facto* contradicts a typical academic definition of plagiarism - *continuous copied words* (CCW) exceeding a given limit. Plagiarism percentage for an uploaded article can also be misleading for cases where only 1% plagiarism is reported if one page in a 100-page article is fully duplicated from the internet, or 50% plagiarism is reported if the second page in a two-page report contains exclusively the references that the software fails to filter out. S.A.V.E. reports both the CCW and the copy percentage of an article and users can rank the examined articles by the longest CCW for each article.

**Similarity and distinction between S.A.V.E. and typical search engines**

Similar to popular search engines nowadays, our PBS-based search also yields search results containing gaps or mismatches when aligning with a query, while our results respect the order of "amino acids" in a PBS more so than keyword-based results. On the other hand, our CCW-based ranking is carried out by prioritizing the highest local similarity, while popular search engines rank the pages containing specific keywords by how "central" these pages are in the network. S.A.V.E. allows the search of the full article (either encoded or plaintext) and a single keyword (plaintext only), while many popular search engines allow <35 keywords without a privacy-protection scheme herein introduced. In this scheme, even the search engine providers cannot know for sure what is searched while users still can get their search results.

# CONCLUSION

S.A.V.E. enabled a new form of search engine where sensitive query content is the outstanding plagiarism detection search engine with the state-of-art bioinformatics algorithm, FM-index. S.A.V.E. has an efficient throughput as Bowtie and 5 order better than BLAST. The encryption method promises confidentiality, where the method has a low false-positive rate when we set the plagiarism threshold at 12 continuous words or above, where text compression is realized by encryption of every word into a single alphabet. S.A.V.E. also recruits a in-house designed reference-removal module, which reaches 0.97 AUC.

# METHODS

**Encryption**

In our current implementation, plaintexts from Microsoft Office documents, Open documents, txt files and pdf files etc., in all languages, are first extracted by tools including Apache Tika. The alphabets in Western languages are recognized as those whose Unicode numbers are less than 1000 and that are not special characters. Any character, e.g., "單" in Chinese (**Figure 3**), "어" in Korean, "ひ" (any hiragana, katakana and kanji) in Japanese, having a Unicode higher than 1000 is considered as an Eastern character/word. Western words are defined as contiguous characters or symbols that have Unicode less than 1000 while being delimited by blanks, tabs, carriage return, '+', '&' and Eastern characters. For example, in the plaintext "單個 pri-mRNA 可以含有 1 至 6 個 miRNA precursor", "pri-mRNA" is considered as one word in English because what are immediately in front and after are Eastern characters "個" and "可", respectively; "miRNA" is another word because the Eastern character/word "個" is in front and one space is trailing behind (**Figure 3**). For a word in Western language being converted into one of the PBS characters ($c$ characters in total), all the Unicode of the constituent characters and symbols in that word are first added up and then divided by $a$. The remainder (0, 1, 2, …or $a$-1) can then be assigned with one of the used PBS characters (see **Figure 3**). Each Eastern character or every symbol is considered as a single word and the single character's Unicode is divided by $c$ to obtain a remainder that is again assigned with a PBS character. By aforementioned approach, any plaintext in any language (or mixed languages) can be converted into a long PBS constituted by a selection of $a$ PBS characters. This irreversible transformation of plaintext runs on users' local machines conferring the foundation of search privacy.

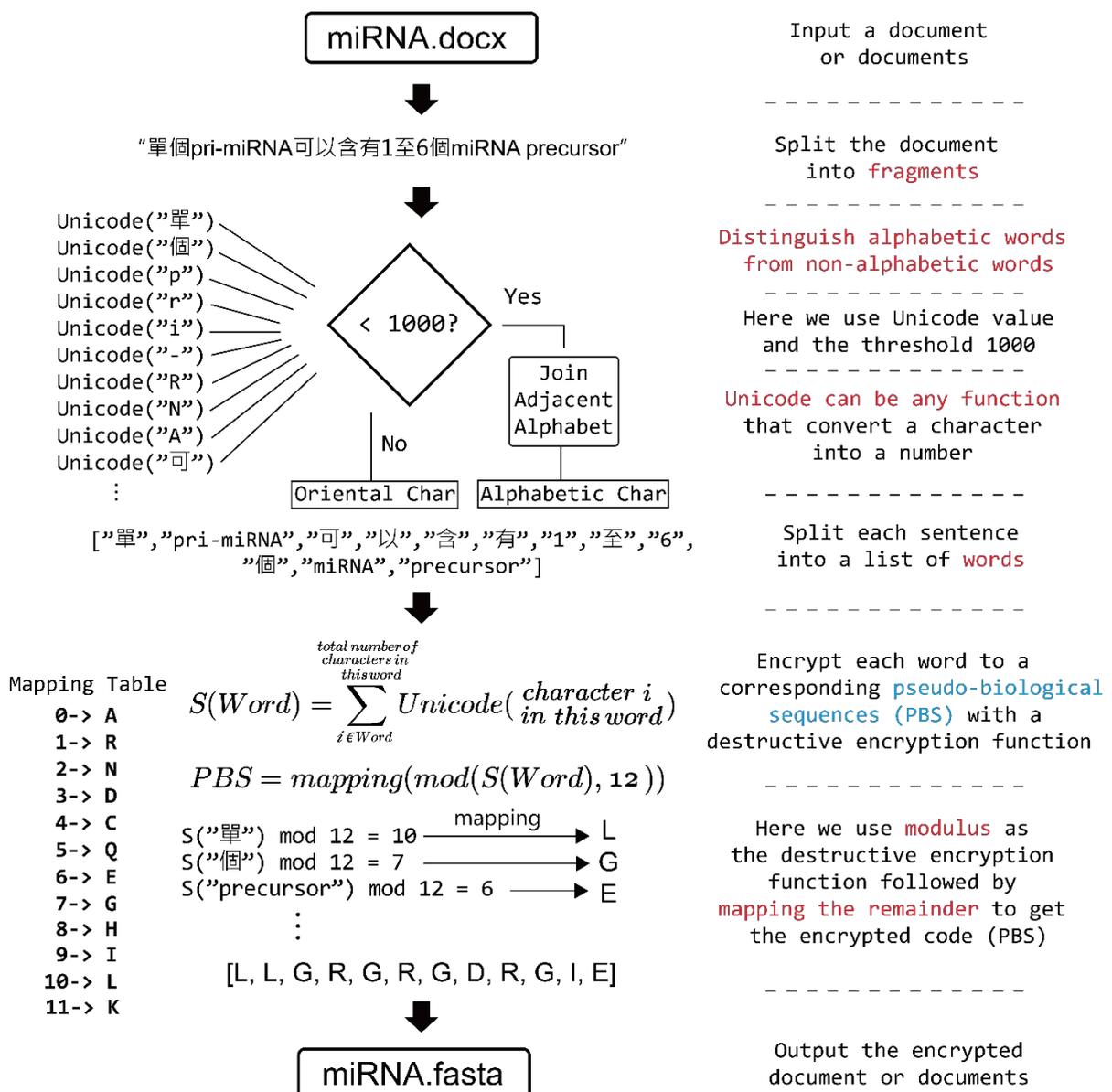

**Figure 3.** An example of the workflow of the encryption process. In this example, the document miRNA.docx contains sentences in mixed languages. All the sentences are extracted by our text parser. Here, we look at a specific text string, "單個 pri-miRNA 可以含有 1 至 6 個 miRNA precursor" where alphabetic words and non-alphabetic words are first identified by whether the Unicode of a certain character is less than 1000. If yes, the characters belong to a Western language; if not, oriental (from

Eastern culture). Those contiguous western characters, delimited by space or a symbol, are considered as a word, based on which encoding is carried out. For example, the first character "單" has a Unicode > 1000, which is deemed as a non-alphabetic (oriental) word, while that of "p", "r", "e", "c", "u", "r", "s", "o", and "r" are less than 1000, which are combined as an alphabetic word (in Western language). To encode a word, the Unicode of all the characters in a given word are first summed and then divided by the total types of PBS characters (here $a$ = 12) to obtain the remainder, followed by converting the remainder to a corresponding "amino acid", a single-letter code in the mapping table. Finally, we output the encrypted file in the FASTA format, which includes a single-line description, followed by a sequence of amino acid characters.

## Search with Full-text index in Minute space (FM-index)

As previously stated, the FM-index is derived from Burrows-Wheeler transform (BWT) and is widely used in various NGS aligners (*16, 17*). FM-index search first converts the raw PBS strings in our database into a sorted suffix array table, where the first column is called *F* and the last column is called *L*, also known as the BWT string (**Figure 4**). First, we need to tally the number of each character in the column *F* and column *L*; for that, we use the C table and occ table to store the tally information for the F and L columns, respectively. The core search method is based on the LF mapping, which allows us to recover the string by mapping the characters of the F column and the L column. For example, suppose we have a query string "DNGQDRGD" (the first *k*-mer in **Figure 5**) and a reference string "EDNGQDRGDQDRN", we want to know where the query string occurs in the reference string (**Figure 5**). The LF mapping searches character by character starting from the back of the string, in this case "D". We look for the position of "D" in the F column by C table, and we know that the range of "D" in the F column is from row 1 to row 4. Next, we search for "G", the next character of "D", in the L column from row 1 to row 4 and in the F column. As Figure 5 shows, the range of "G" in the F column is from row 6 to row 7 and row 2 has "D" in the F column and "G" in the L column, where the "G" is the first "G" in the L column (the second "G" is at row 10). Accordingly, we find that the first "G" in the L column is the same "G" in the F column. Hence, we find that the prefix of row 6 contains "GD", which is the last two characters of our query. The LF mapping follows the same procedures above, keep searching for "R", "D", "Q", "G", "N", and "D". Finally, we can find that the query string "DNGQDRGD" is at row 1 in the suffix array table, which can then be converted back to the real position in the reference string. The time complexity of FM-index search is O(m), where m is the length of the query string. Since m is normally shorter than the length of the database, the FM-index search is the most efficient algorithm known to such use cases (*14*).

## Plagiarism Detection

Plagiarism detection in S.A.V.E. can be any method with the sequence comparison ability, especially widely used sequencing alignment methods in bioinformatics. All text converted into PBS in the encryption step can directly fit into those algorithms. In the implementation of the current version of S.A.V.E., we used BWT (*15*) and FM index (*14*) for plagiarism detection, which is also a widely used algorithm in sequencing alignment of bioinformatics.

The alignment algorithms with BWT and FM index can efficiently search the exact match between query and reference. To apply BWT and FM index in plagiarism detection, we first need to build FM-index, the fastest among its kind, for references as the database. After establishing the database, we can perform plagiarism detection on the encrypted content efficiently due to the reduced time complexity of the algorithm. Furthermore, as we have encrypted all plaintext into PBS, the original content of the query document can never be exposed during the searching process.

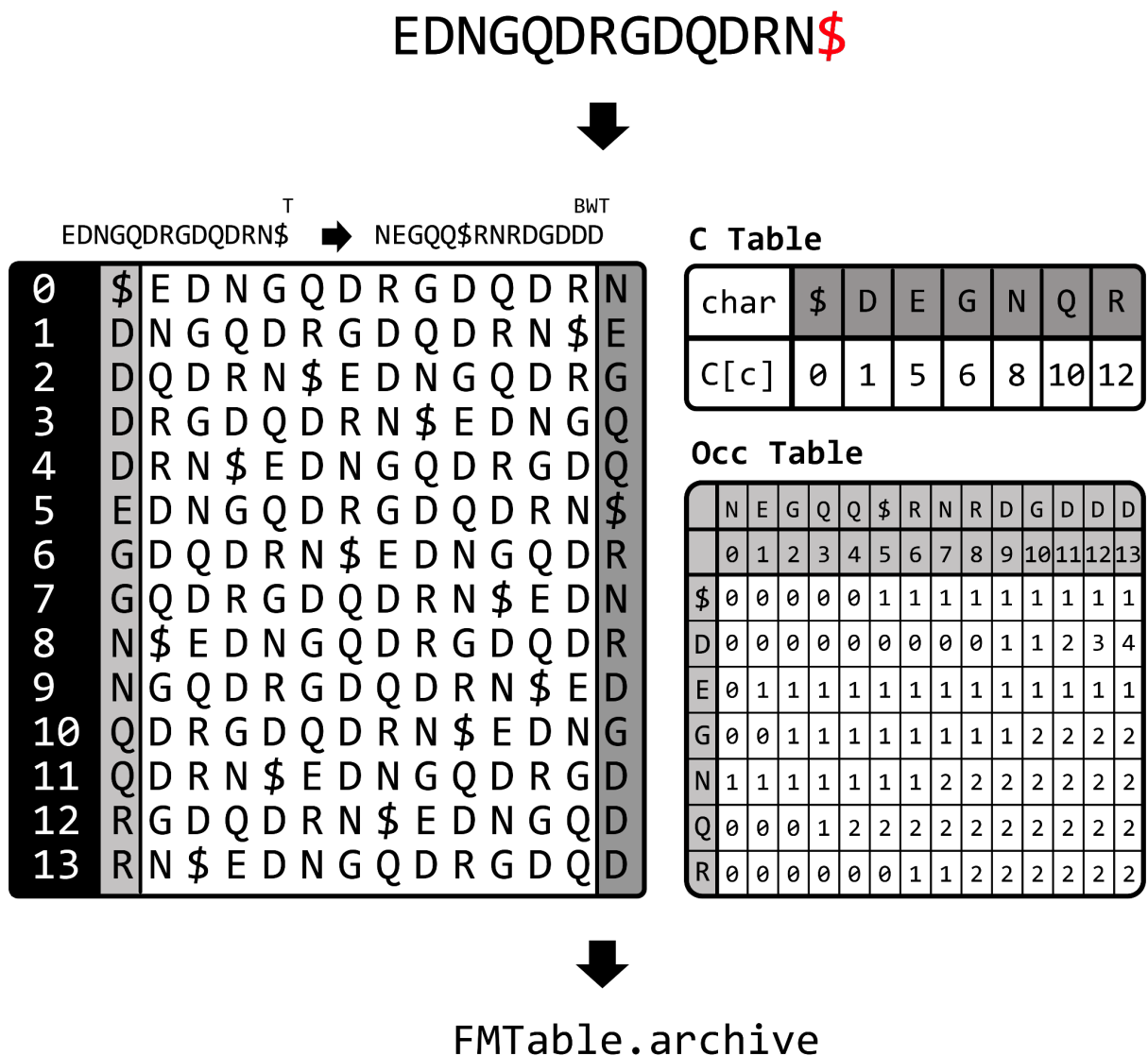

**Figure 4.** An example of constructing an FM-index for reference as a database. To construct a database for plagiarism detection, we need to concatenate the encrypted reference strings and add a dollar sign at the end of the joint string followed by Burrows-Wheeler Transformation to form a BWT string and create a sorted suffix array matrix, a character table (c table) that counts the total number for each character, and occ table that counts the accumulated number of each character at a specific position of BWT string. After that, we archived those tables as the database for plagiarism detection.

**Build Database**

In order to perform the plagiarism detection on PBS generated from query documents, the database, or pre-built FM-index table, is necessary. In this study, we constructed databases with English Wikipedia pages, Chinese Wikipedia pages, PubMed Open Access articles, and PubMed Baseline articles as of end of 2021, where raw data are in XML. Therefore, we parsed and extracted text from these data followed by converting the content into PBS and formatting it as FASTA using the same destructive encryption method mentioned above. The table is constructed from these FASTA files, including three parts, sorted suffix array table, character table (C table), and cumulative occurrence table (Occ table), which are all indispensable for query searching (**Figure 4**). To date, we have collected 146.1GB of text data with total 17,802,587 Wikipedia pages, 5,613,955 PubMed Open Access articles, and 40,150,958 abstracts.

**Pairwise comparison and Temporary Database**

For submissions of a set of encrypted documents, S.A.V.E. performs plagiarism detection on these documents, checking plagiarism content among the documents, called pairwise comparison (**Figure 1A**). Since the current implementation of S.A.V.E. plagiarism detection is based on FM-index, S.A.V.E. builds an FM-index with the documents before performing plagiarism detection, where the FM-index is used for this particular set of documents only.

**Search Against Database**

As mentioned, we used an FM-index-based searching algorithm for plagiarism detection. In detail, the encrypted document, where plaintext has been converted into ciphertext (PBS), will be split into $k$-mers (default $k$=8). And each $k$-mer will search reversely from the end of the $k$-mer for exact matches on the database. The results will then be merged and report all matched sequences and their position aligned to the database. An example of the protection search was shown in Figure 5.

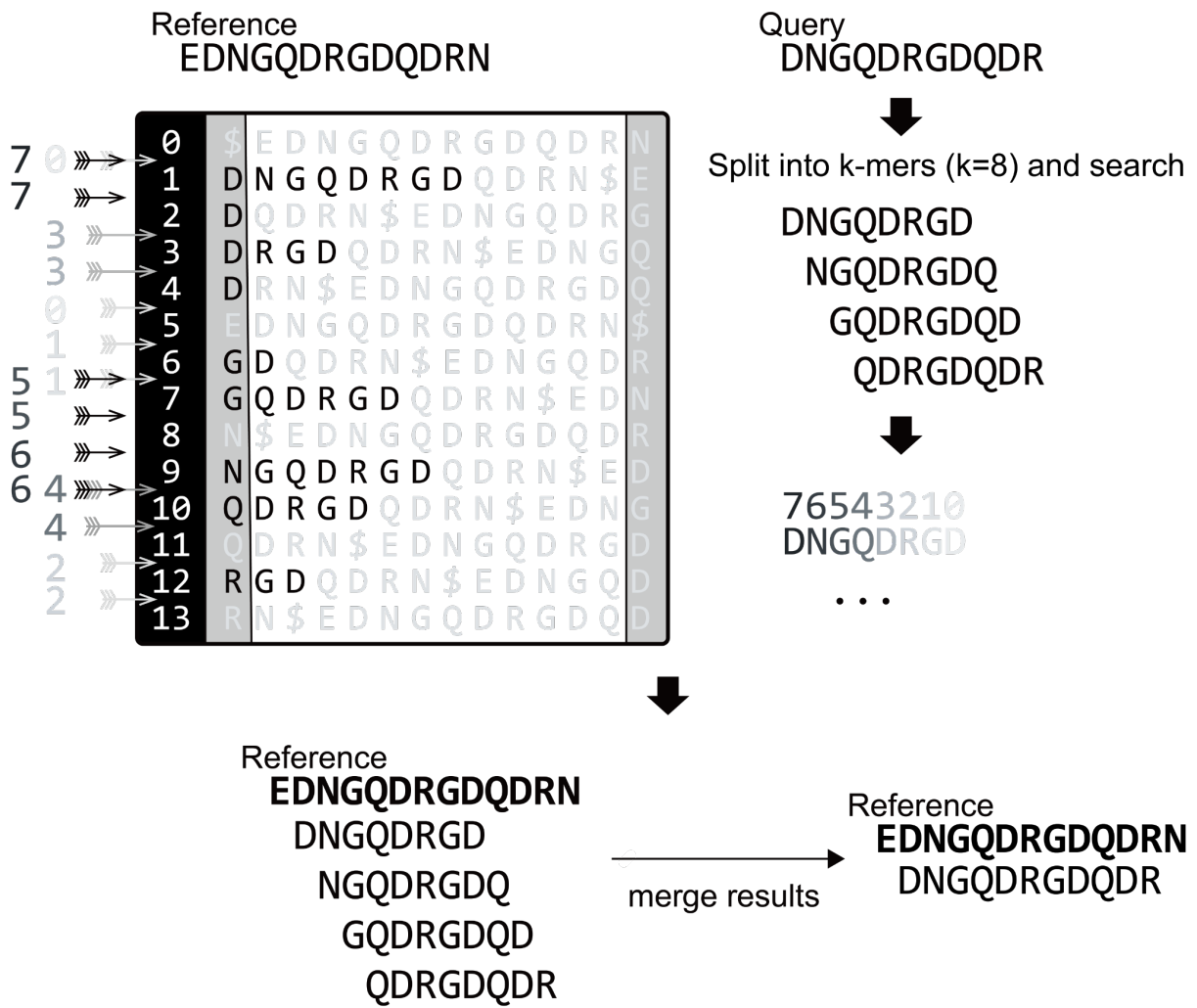

**Figure 5.** An alignment procedure example. The alignment procedure is the core of our plagiarism detection, which is not a typical string-matching algorithm. Before the alignment procedure begins, we will load the reference string that has been converted to archived FM-index tables (see **Figure 4**), including a suffix array matrix shown on the left. A query string is first split into *k*-mers (*k*=8, in this example), while each *k*-mer is searched against the FM-index tables, base by base, where the searching order is in grayscale from light to dark (or index 0 to 7). The searching results of each step with its range on the suffix array matrix are displayed as arrows with the same grayscale color and index on the left as well. After searching through all *k*-mers, we then merge the results reporting the aggregated result.

**The Metrics of Plagiarism**

In order to make it easier for users to identify if a given document contains plagiarized texts, we quantify the extent of plagiarism by reporting a whole-document similarity percentage (copy coverage) as well as the number of continuously copied words (CCW) for each potentially plagiarized paragraph. Herein, CCW score is a better measure of plagiarism than the similarity score because it follows a well-accepted academic standard to define plagiarism. CCW score is the number of consecutive words in the queried document that are found identical to those on the internet or in other documents uploaded together with this query file. In such a string-matching procedure, we allow groups of mismatches up to three consecutive words.

**The Logistic Regression for Detecting Reference**

The reference detector, a logistic regression model, takes the text in an article as the input and goes through it line by line to mark each line as either reference or text. **Table 3** shows the list of patterns that commonly appear in the reference. The algorithm counts the appearance of the patterns in each line, normalized by the total number of characters in that line.

After scanning through the entire document, a pattern density matrix, **D**, in a dimension of $N_{line}$ by $N_{pattern}$ is established, where $N_{pattern}$ is the number of patterns (currently 19, listed in **Table 3**) we examined for a given line, and $N_{line}$ represents the total number of lines in the document. As the references are usually listed together (usually toward the end), we apply the moving average of window with a size of 3 to allow the pattern density per line to represent a neighborhood of 3 lines. The moving average of line density is written into the matrix $\underline{\mathbf{D}}$ of the same size of **D**. The logistic regression model is listed below.

$$P_{ref}(z_i) = \frac{1}{(1 - e^{-z_i})}$$

$$z_i = \underline{\mathbf{D}}_i \mathbf{W}^T + c$$

where $i$ indexes the line; **W** is a row vector of weight parameters, each of which is assigned for the on and off of one pattern and $c$ is a constant. $P_{ref}$ is a probability of line $i$ of the document being a reference. The parameters **W** listed in **Table 3** are optimized by Maximum Likelihood Estimation (*22*) using 100 journal papers and 200 student assignments as the training examples where each line is manually assigned as reference or non-reference text.

**Table 3. The list of the common patterns used in the reference detection**

| Description of the pattern | Pattern in regular expression | Weight |
|---|---|---|
| Semicolon | ; | 15.78814 |
| A capital alphabet followed by a full-stop and a comma (e.g. J.,) | [A-Z]\., | 12.57611 |
| Asterisk | \* | 11.6191 |
| Ampersand | \& | 11.45578 |
| 4 digits enclosed in a bracket | \(\d{4}\) | 10.6957 |
| An upper-case alphabet followed by a full-stop, flanked by a blank space each side. | (?<=\s)[A-Z]\.(?=\s) | 9.418422 |
| Identifies the first or middle initials of an author. | | |
| Identifies the journal volume and page numbers separated by a colon | \d{0,1}:\s*\d+[--]\d+ | 9.050281 |
| Identifies a year | (?<=\s)\d{4}[,\.] | 7.692744 |
| Three digits at the beginning of each line. (The index of biography) | | 6.614513 |
| Identifies the journal volume and page numbers separated by a comma or full-stop. | [,\.]\s*\d+[--]\d+ | 6.412431 |
| A blank space followed by an upper-case alphabet and a comma. Identifies the first or middle initials of an author. | \s[A-Z], | 5.971774 |
| Identifies the first and middle initials of an author. (e.g. Chang **Y.Y.** ) | (?<=\s)[A-Z]\.[A-Z]\.(?=\s) | 5.59754 |
| DOI | (?<!\w)doi\|DOI(?!\w) | 5.075362 |
| et al | et al | 2.745097 |
| Identifies names | (?<=\s)[A-Z]\w+(?=\s) | -0.35329 |
| , | , | -0.71845 |
| A capital alphabet | [A-Z] | -0.81583 |
| Identifies names followed by a comma | \s[A-Z][a-z]+, | -0.82969 |
| Identifies the index of a reference | \[.*\] | -1.9572 |

## Acknowledgements

We acknowledge the programming support and advice from Drs Soumya Ray, Chi-Ching Lee and Hongchun Lee as well as Mr Chia-Yung Shih. E.O.S. acknowledges financial support from Taiwan International Graduate Program, Academia Sinica, Taipei, Taiwan. Part of this work was financially supported by the National Tsing Hua University, Taiwan (10020B002).

## References

1.   O. Tene, What Google Knows: Privacy and Internet Search Engines. *SSRN Electronic Journal*,  (2007).

2. C. Chiru, Search engines : ethical implications. Economics, management and financial markets. **11**, p. 162-167 (2016).

3. A. Hands, Duckduckgo http://www.duckduckgo.com or http://www.ddg.gg. *TECHNICAL SERVICES QUARTERLY* **29**, 345-347 (2012).

4. J. H. Allen, Ashley Privacy and Security Tips for Avoiding Financial Chaos. *Aspen Publishers*, 101-107 (2018).

5. R. Ridgway, Against a Personalisation of the Self. *Ephemera: Theory & politics in organization* **17**, 377-397 (2017).

6. A. D. Born, Teaching Tip: How to Reduce Plagiarism. *Journal of Information Systems Education* **14**, 223-224 (2003).

7. M. J. Austin, L. D. Brown, Internet Plagiarism: Developing Strategies to Curb Student Academic Dishonesty. *The Internet and Higher Education* **2**, 21-33 (1999).

8. Y. Kauffman, M. F. Young, Digital plagiarism: An experimental study of the effect of instructional goals and copy-and-paste affordance. *Computers & Education* **83**, 44-56 (2015).

9. K. Vani, D. Gupta, Study on extrinsic text plagiarism detection techniques and tools. *Journal of Engineering Science and Technology Review* **9**, 150-164 (2016).

10. K. B. H. Kanakappa I Narasanaikar, Plagiarism: An electronic detection tool Turnitin. *International Journal of Multidisciplinary Research and Development* **4**, 187-190 (2017).

11. M. Errami, J. D. Wren, J. M. Hicks, H. R. Garner, eTBLAST: a web server to identify expert reviewers, appropriate journals and similar publications. *Nucleic Acids Research* **35**, W12-W15 (2007).

12. S. Niezgoda, T. Way, *SNITCH: A software tool for detecting cut and paste plagiarism.* (2006), vol. 38, pp. 51-55.

13. C. Hilton, *United States District Court for the Eastern District of Virginia, Alexandria Division* **07-0293**, (2008).

14. P. Ferragina, G. Manzini, in *Proceedings 41st Annual Symposium on Foundations of Computer Science.* (2000), pp. 390-398.

15. M. B. D. J. Wheeler, A Block-Sorting Lossless Data Compression Algorithm. *Technical Report* **124**, (1994).

16. B. Langmead, C. Trapnell, M. Pop, S. L. Salzberg, Ultrafast and memory-efficient alignment of short DNA sequences to the human genome. *Genome Biol* **10**, R25 (2009).

17. H. Li, R. Durbin, Fast and accurate short read alignment with Burrows-Wheeler transform. *Bioinformatics* **25**, 1754-1760 (2009).

18. R. Li *et al.*, SOAP2: an improved ultrafast tool for short read alignment. *Bioinformatics* **25**, 1966-1967 (2009).

19. S. F. Altschul, W. Gish, W. Miller, E. W. Myers, D. J. Lipman, Basic local alignment search tool. *J Mol Biol* **215**, 403-410 (1990).

20. B. Arnoldy, Students sue antiplagiarism website for rights to their homework. *The Christian Science Monitor*, (2007).

21. C.-H. Chang *et al.*, sBWT: memory efficient implementation of the hardware-acceleration-friendly Schindler transform for the fast biological sequence mapping. *Bioinformatics* **32**, 3498-3500 (2016).

22. S. Menard, *Applied Logistic Regression Analysis.* (SAGE Publications, 2002).